\documentclass{elsart}
\begin{document}
\date{24 January 2005}
\begin{frontmatter}
\title{Exactly solvable nonlinear model with two multiplicative
Gaussian colored noises}
\author{A.N. Vitrenko}
\ead{vitrenko\_andrei@mail.ru}
\address{Department of Mechanics and Mathematics, Sumy State
University, 2, Rimskiy-Korsakov Street, 40007 Sumy, Ukraine}

\begin{abstract}
An overdamped system with a linear restoring force and two
multiplicative colored noises is considered. Noise amplitudes
depend on the system state $x$ as $x$ and $|x|^{\alpha}$. An
exactly soluble model of a system is constructed due to
consideration of a specific relation between noises. Exact
expressions for the time-dependent univariate probability
distribution function and the fractional moments are derived.
Their long-time asymptotic behavior is investigated analytically.
It is shown that anomalous diffusion and stochastic localization
of particles, not subjected to a restoring force, can occur.
\begin{keyword}
colored noises \sep Gaussian processes \sep statistical properties
\sep anomalous diffusion
\PACS{05.40.-a}
\end{keyword}
\end{abstract}
\journal{Physica A}

\end{frontmatter}
\newpage

\section{Introduction}

A time evolution of a stochastic system can be described by a
Langevin equation \cite{Kampen}. In such approach, the influence
of a fluctuating environment is accounted for by means of an
external noise term. It is necessary to determine exact
statistical characteristics of a system in terms of given
statistical characteristics of a noise. Assumption of Gaussian
white (delta-correlated) noise considerably simplifies the problem
\cite{Horsthemke}. The time evolution of the system is Markovian
process \cite{Doob}, and its univariate probability distribution
function (PDF) and transition probability density satisfy the
Fokker-Planck equation \cite{Horsthemke,Gardiner,Risken}. In
specific cases one can determine an exact time-dependent solution
of such an equation, and a stationary PDF can be obtained in
general.

However, Gaussian white noise has some unphysical properties
\cite{HanggiJung}, therefore sometimes its application is not
justified. Gaussian colored noise with an arbitrary correlation
function is more realistic model of a fluctuating environment.
Although the time evolution of the stochastic system is
non-Markovian process, in specific cases its univariate PDF
satisfies the time dependent Fokker-Planck equation
\cite{Hanggi78,Hanggi81,Jung,DH65}. Exact statistical
characteristics of undamped \cite{MW} and damped free particles
\cite{DH62} driven by additive Gaussian colored noise of
overdamped particles in a quadratic potential driven by
multiplicative Gaussian colored noise \cite{DH65} are found. In
particular, it is shown that anomalous diffusion
\cite{MW,DH62,DH65} and stochastic localization \cite{DH62,DH65}
of free particles can occur.

In a more general case, a fluctuating environment of a system is
modeled by two noise sources. Most commonly, cross-correlated
Gaussian white noises are considered. The appropriate
Fokker-Planck equation is derived and the two-noise Langevin
equation can be reduced to a stochastically equivalent one-noise
Langevin equation \cite{FT,WCK,DVH}. However, models which
demonstrate, for example, nonequilibrium fluctuation-induced
transport \cite{Magnasco,DHR,BRH}, double stochastic resonance
\cite{JZHL}, require to consider a colored noise source and a
white noise source. The Ornstein-Uhlenbeck process with
exponential correlation function is widely used as Gaussian
colored noise. But the corresponding Fokker-Planck equation is
derived using approximate methods, for instance, conventional
small-$\tau$ theory \cite{SMKG}, the decoupling theory
\cite{HMMM}, the unified colored noise approximation \cite{JH},
etc. Within this approximations, systems driven by two
Ornstein-Uhlenbeck processes are studied \cite{WCW,LCWW}, the
noise self-correlations and the cross-correlation have the same
correlation time. There are no general methods for obtaining exact
statistical characteristics of systems with two colored noises.
Therefore, specific exactly soluble models are needed.

In this paper, we generalize a nonlinear system \cite{DH65} with
colored noise to the case of two colored noises and study the
possibility of an exactly solvable model constructing. The system
state parameter $x(t)$ evolves according to the following Langevin
equation:
\begin{equation}
    \dot{x}(t)+[\kappa+f_{1}(t)]x(t)=|x(t)|^{\alpha}f_{2}(t),
    \label{eq:Lang_x}
\end{equation}
where $x(0)=x_{0}(>0)$, $\kappa (\geq 0)$ and $\alpha (<1)$ are
real-valued parameters, $f_{1}(t)$ and $f_{2}(t)$ are colored
noises with zero mean and known statistical characteristics. The
assumption of positivity of $x(0)$ does not restrict the
generality of the model (\ref{eq:Lang_x}).

We note that for $0\leq\alpha\leq1$ the solution of
Eq.~(\ref{eq:Lang_x}) is not unique at $x=0$. In this paper, the
point $x=0$ is considered to be a regular point and the solution
of Eq.~(\ref{eq:Lang_x}) coincides with the solution of the
equation
\begin{equation}
    \{\dot{x}(t)+[\kappa+f_{1}(t)]x(t)\}|x(t)|^{-\alpha}=f_{2}(t)
    \quad [x(0)=x_{0}>0]
    \label{eq:Lang_x1}
\end{equation}
for all times $t\geq 0$. Although one can obtain an exact solution
of Eq.~(\ref{eq:Lang_x1}) for the case of arbitrary independent
noises $f_{1}(t)$ and $f_{2}(t)$, it is not possible to determine
exact statistical characteristics of $x(t)$. Therefore, we
consider a specific relation between noises $f_{1}(t)$ and
$f_{2}(t)$, which allows to obtain exactly the time-dependent
univariate PDF and the fractional moments of $x(t)$. We find
analytically their long-time asymptotics and show that in the
special case $\kappa=0$ anomalous diffusion and stochastic
localization of particles can occur.

The paper is organized as follows. In Sec.\ II, we solve exactly
Eq.~(\ref{eq:Lang_x1}) with arbitrary independent noises
$f_{1}(t)$, $f_{2}(t)$ and consider a specific relation between
noises. In the same section we obtain the time-dependent
univariate PDF and the fractional moments of $x(t)$. Their
long-time asymptotic behavior is studied analytically in Sec.\
III. Our results are summarized in Sec.\ IV.

\section{General analysis}

To construct an exactly soluble model of a system described by the
Langevin equation (\ref{eq:Lang_x1}), we obtain its exact solution
for the case of arbitrary independent noises $f_{1}(t)$ and
$f_{2}(t)$. By introducing the new variable
$y(t)=x(t)|x(t)|^{-\alpha}$, the nonlinear stochastic differential
equation (\ref{eq:Lang_x1}) is transformed to the linear one
\begin{equation}
    \dot{y}(t)+[\omega+(1-\alpha)f_{1}(t)]y(t)=(1-\alpha)f_{2}(t),
    \label{eq:Lang_y}
\end{equation}
where $\omega=(1-\alpha)\kappa$ and $y(0)=x_{0}^{1-\alpha}$. The
solution of Eq.~(\ref{eq:Lang_y}) is given by (see, for example,
Ref. \cite{Kamke})
\begin{equation}
    y(t)=e^{-\omega t-(1-\alpha)F_{1}(t)}\left[x_{0}^{1-\alpha}+
    (1-\alpha)\int_{0}^{t}d\tau f_{2}(\tau)e^{\omega\tau+(1-\alpha)
    F_{1}(\tau)}\right],
    \label{eq:Sol_y}
\end{equation}
where
\begin{equation}
    F_{1}(t)=\int_{0}^{t}d\tau f_{1}(\tau).
    \label{eq:F1_t}
\end{equation}
Substituting Eq.~(\ref{eq:Sol_y}) into expression
$x(t)=y(t)|y(t)|^{\alpha/(1-\alpha)}$, we can write the solution
of Eq.~(\ref{eq:Lang_x1}) as
\begin{equation}
    x(t)=\frac{F(t)|F(t)|^{\alpha/(1-\alpha)}}{e^{F_{1}(t)}},
    \label{eq:Sol_x}
\end{equation}
where
\begin{equation}
    F(t)=e^{-\omega t}\left[x_{0}^{1-\alpha}+(1-\alpha)
    \int_{0}^{t}d\tau
    f_{2}(\tau)e^{\omega\tau+(1-\alpha)F_{1}(\tau)}\right].
    \label{eq:F_t}
\end{equation}

To obtain the time-dependent univariate PDF of $x(t)$, we should
determine the univariate PDFs of $F_{1}(t)$ and $F(t)$. The
univariate PDF of $F_{1}(t)$ is easily found if $f_{1}(t)$ is
Gaussian noise with zero mean and the arbitrary correlation
function
\begin{equation}
    \langle f_{1}(t)f_{1}(t')\rangle=r_{1}(|t-t'|).
    \label{eq:corr_fun_f1}
\end{equation}
Indeed, by virtue of the linear dependence $F_{1}(t)$ on
$f_{1}(t)$, $F_{1}(t)$ is Gaussian process with zero mean and the
dispersion
\begin{equation}
    \sigma_{1}^{2}(t)=2\int_{0}^{t}du r_{1}(u)(t-u).
    \label{eq:disp1}
\end{equation}
The simplest way to obtain the univariate PDF of $F(t)$ is to
consider the specific relation between noises $f_{1}(t)$ and
$f_{2}(t)$. We assume that
\begin{equation}
    f_{2}(t)=f(t)e^{(\alpha-1)F_{1}(t)},
    \label{eq:intr_f}
\end{equation}
where $f(t)$ is Gaussian noise with zero mean and the arbitrary
correlation function
\begin{equation}
    \langle f(t)f(t')\rangle=r(|t-t'|).
    \label{eq:corr_fun_f}
\end{equation}
Then $F(t)$ is Gaussian noise with the mean $m(t)$ and the
dispersion $\sigma^{2}(t)$
\begin{equation}
    m(t)=x_{0}^{1-\alpha}e^{-\omega t},
    \label{eq:mean}
\end{equation}
\begin{equation}
    \sigma^{2}(t)=2(1-\alpha)\frac{e^{-\omega t}}{\kappa}
    \int_{0}^{t}du r(u)\mathrm{sinh}[\omega(t-u)].
    \label{eq:disp}
\end{equation}
Thus, the Langevin equation (\ref{eq:Lang_x1}) with arbitrary
noises $f_{1}(t)$ and $f_{2}(t)$ is reduced to the following
Langevin equation:
\begin{equation}
    \{\dot{x}(t)+[\kappa+f_{1}(t)]x(t)\}|x(t)|^{-\alpha}=
    f(t)e^{(\alpha-1)F_{1}(t)} \quad [x(0)=x_{0}>0],
    \label{eq:Lang_x2}
\end{equation}
where $f_{1}(t)$ and $f(t)$ are statistically independent Gaussian
colored noises with known statistical characteristics.

The random process $p(t)=F(t)|F(t)|^{\alpha/(1-\alpha)}$ is
power-normal \cite{DH65}. Let $P_{p}(p,t)$ denote the univariate
PDF that $p(t)=p$,
\begin{equation}
    P_{p}(p,t)=\frac{1-\alpha}{\sqrt{2\pi}\sigma(t)|p|^{\alpha}}
    \exp\left\{-\frac{[p|p|^{-\alpha}-m(t)]^{2}}
    {2\sigma^{2}(t)}\right\}.
    \label{eq:Pp1}
\end{equation}
The random process $l(t)=e^{F_{1}(t)}$ is logarithmic-normal
\cite{LN}. Let $P_{l}(l,t)$ denote the univariate PDF that
$l(t)=l$,
\begin{equation}
    P_{l}(l,t)=\frac{1}{\sqrt{2\pi}\sigma_{1}(t)l}\exp\left[
     -\frac{\ln^{2}l}{2\sigma_{1}^{2}(t)}\right]
    \label{eq:Pl1}
\end{equation}
for $l\geq 0$ and $P_{l}(l,t)=0$ for $l<0$. According to
Eq.~(\ref{eq:Sol_x}), the solution of Eq.~(\ref{eq:Lang_x2}) can
be rewritten as $x(t)=p(t)/l(t)$, where $p(t)$ and $l(t)$ are
statistically independent power-normal and lognormal processes. It
is not difficult to get the conclusion, that by introducing the
new variable $p(t)=x(t)l(t)$, Eq.~(\ref{eq:Lang_x2}) can be
obtained from the Langevin equation with one noise
\begin{equation}
    [\dot{p}(t)+\kappa p(t)]|p(t)|^{-\alpha}=f(t)\quad
    [p(0)=p_{0}>0],
    \label{eq:Lang_P}
\end{equation}
where $p_{0}=x_{0}$.

Let $P_{x}(x,t)$ be the univariate PDF that $x(t)=x$. It can be
calculated as (see, for example, Ref. \cite{Papoulis})
\begin{equation}
    P_{x}(x,t)=\int_{0}^{\infty}duuP_{l}(u,t)P_{p}(ux,t).
    \label{eq:Px}
\end{equation}
Substituting Eq.~(\ref{eq:Pp1}) and Eq.~(\ref{eq:Pl1}) into
Eq.~(\ref{eq:Px}), we finally obtain the time-dependent univariate
PDF of $x(t)$
\begin{equation}
    P_{x}(x,t)=\frac{1-\alpha}{2\pi\sigma_{\alpha}(t)\sigma(t)
    |x|^{\alpha}}\int_{0}^{\infty}du\exp\left\{-\frac{\ln^{2}u}
    {2\sigma_{\alpha}^{2}(t)}-\frac{[ux|x|^{-\alpha}-m(t)]^{2}}
    {2\sigma^{2}(t)}\right\},
    \label{eq:Px1}
\end{equation}
where $\sigma_{\alpha}(t)=(1-\alpha)\sigma_{1}(t)$.

We note if $f_{1}(t)\equiv 0$ [$\sigma_{\alpha}^{2}(t)\rightarrow
0$ for all times $t$] the small neighborhood of the point $u=1$
gives basic contribution and Eq.~(\ref{eq:Px1}) is reduced to the
PDF of the power-normal distribution (\ref{eq:Pp1}). If
$f(t)\equiv 0$ [$\sigma^{2}(t)\rightarrow 0$ for all times $t$] we
take into account
\begin{eqnarray}
    \lim_{\sigma^{2}(t)\rightarrow 0}\Bigg\{\left\{\sqrt{2\pi}
    \left[\frac{\sigma(t)}{|x|^{1-\alpha}}\right]\right\}^{-1}
    \exp\Bigg\{-\frac{1}{2}\left[u-\frac{m(t)}{x|x|^{-\alpha}}
    \right]^2\nonumber\\
    \times\left[\frac{\sigma(t)}{|x|^{1-\alpha}}\right]^{-2}
    \Bigg\}\Bigg\}=\delta\left[u-\frac{m(t)}{x|x|^{-\alpha}}\right],
    \qquad\quad
    \label{eq:delta}
\end{eqnarray}
where $\delta(x)$ is the Dirac delta function. Eq.~(\ref{eq:Px1})
is reduced to the PDF of the lognormal distribution \cite{DH65}
\begin{equation}
    P_{x}(x,t)=\frac{1}{\sqrt{2\pi}\sigma_{1}(t)x}
    \exp\left[-\frac{1}{2\sigma_{1}^{2}(t)}\left(\ln\frac{x}{x_{0}}+
    \kappa t\right)^{2}\right]
    \label{eq:Px0}
\end{equation}
for $x\geq 0$ and $P_{x}(x,t)=0$ for $x<0$. Taking into account
Eqs.~(\ref{eq:delta}) and (\ref{eq:Px0}), we note that
Eq.~(\ref{eq:Px1}) reduces to the delta function $\delta(x-x_0)$
at time $t=0$.

We can now determine the time-dependent fractional moments of
$x(t)$. They are defined as
\begin{equation}
    m_{r}^{v}(t)=\int_{-\infty}^{\infty}dxP_{x}(x,t)|x|^{r-v}x^v,
    \label{eq:frac_mon_def1}
\end{equation}
where $r$ is a real number, and $v=0$ or 1. We substitute
Eq.~(\ref{eq:Px}) into Eq.~(\ref{eq:frac_mon_def1}). By
introducing the new variable $y=ux$, one can write
\begin{equation}
    m_{r}^{v}(t)=\int_{0}^{\infty}\frac{du}{u^{r}}P_{l}(u,t)
    \int_{-\infty}^{\infty}dyP_{p}(y,t)|y|^{r-v}y^{v}.
    \label{eq:frac_mom2}
\end{equation}
We take into account that the second integral is the well-known
fractional moments of power-normal distribution \cite{DH65}.
Calculating the first integral, we finally obtain the fractional
moments of $x(t)$
\begin{eqnarray}
    m_{r}^{v}(t)&=&\frac{\Gamma(\xi)}{\sqrt{2\pi}}\sigma^{\xi-1}(t)
    \exp\left[r^2\sigma_{1}^2(t)/2-a^2(t)/4\right] \nonumber\\
    &&\times\{D_{-\xi}[-a(t)]+(-1)^vD_{-\xi}[a(t)]\}.
    \label{eq:frac_mom}
\end{eqnarray}
Here $\xi=1+r/(1-\alpha)$, $a(t)=m(t)/\sigma(t)$, and
$D_{-\xi}(z)$ is the integral representation of the Weber
parabolic cylinder function \cite{WPCF}
\begin{equation}
    D_{-\xi}(z)=\frac{e^{-z^2/4}}{\Gamma(\xi)}\int_{0}^{\infty}
    dyy^{\xi-1}e^{-y^2/2-zy} \;(\xi>0),
    \label{eq:WPCF}
\end{equation}
where $\Gamma(\xi)$ is the gamma function. If $\xi\leq 0$ all
fractional moments diverge. Since $m_{0}^{0}(t)=1$, $P_{x}(x,t)$
is properly normalized.

We note the asymptotic of $m_{r}^{v}(t)$ as
$\sigma_{1}^{2}(t)\rightarrow 0$ for all times $t$ is obtained
directly from Eq.~(\ref{eq:frac_mom})
\begin{equation}
    m_{r}^{v}(t)=\frac{\Gamma(\xi)}{\sqrt{2\pi}}e^{-a^2(t)/4}
    \sigma^{\xi-1}(t)\{D_{-\xi}[-a(t)]
    +(-1)^vD_{-\xi}[a(t)]\}.
    \label{eq:fm_asymp1}
\end{equation}
Eq.~(\ref{eq:fm_asymp1}) is the fractional moments of the
power-normal distribution \cite{DH65}. In order to determine the
fractional moments as $\sigma^{2}(t)\rightarrow 0$ for all times
$t$, we use the asymptotic form
\begin{eqnarray}
    \lim_{\sigma^{2}(t)\rightarrow 0} \sigma^{\xi-1}(t)
    \exp\left[-\frac{m^{2}(t)}{4\sigma^{2}(t)}\right]
    D_{-\xi}\left[\pm\frac{m(t)}{\sigma(t)}\right]    \nonumber\\
    =\frac{\sqrt{2\pi}}{\Gamma(\xi)}m^{\xi-1}(t)\int_{0}^{\infty}
    dss^{\xi-1}\delta(s\pm1).
    \label{eq:asymp}
\end{eqnarray}
Substituting Eq.~(\ref{eq:asymp}) into Eq.~(\ref{eq:frac_mom}), we
obtain
\begin{equation}
    m_{r}^{v}(t)=x_{0}^{r}\exp\left[\frac{1}{2}r^2
    \sigma_{1}^2(t)-r\kappa t\right]
    \label{eq:fm_asymp2}.
\end{equation}
Eq.~(\ref{eq:fm_asymp2}) is the fractional moments of the
logarithmic-normal distribution \cite{DH65}.

\section{Asymptotic behavior}

In the previous section, we have constructed an exactly soluble
model with two multiplicative Gaussian colored noises. We have
obtained exact expressions for the time-dependent univariate
probability distribution function and the fractional moments. In
this section, we determine the long-time behavior of particles,
$t\rightarrow \infty$. Cases $\kappa>0$ and $\kappa=0$ are
considered separately.

\subsection{$\kappa>0$}

In this case $m(\infty)=0$ and $a(\infty)=0$. The asymptotic form
of Eq.~(\ref{eq:disp}) at $t=\infty$ is given by
\begin{equation}
    \sigma^2(\infty)=\frac{1-\alpha}{\kappa}\int_{0}^{\infty}
    dur(u)e^{-\omega u}.
    \label{eq:disp_inf}
\end{equation}
Since $r(u)\rightarrow 0$ as $u\rightarrow \infty$,
$\sigma^2(\infty)<\infty$. The asymptotic form of
Eq.~(\ref{eq:disp1}) can be written as
\begin{equation}
    \sigma_{1}^{2}(t)\sim 2tR_{1}(t)\;(t\rightarrow\infty),
    \label{eq:disp1_inf}
\end{equation}
where $R_{1}(t)=\int_{0}^{t}du r_{1}(u)$. According to Ref.
\cite{DH62}, if
\begin{equation}
    R_{1}(t)=o(1/t)\;(t\rightarrow \infty),
    \label{eq:R1}
\end{equation}
then $\sigma_{1}^2(\infty)<\infty$, and if $0<R_{1}(\infty)\leq
\infty$ or $R_{1}(\infty)=0$ [$R_{1}(\infty)$ is the noise
intensity], but Eq.~(\ref{eq:R1}) does not hold, then
$\sigma_{1}^2(\infty)=\infty$. Thus, there are two qualitatively
different cases of the long-time asymptotic behavior of
Eq.~(\ref{eq:disp1}), namely, $\sigma_{1}^2(\infty)<\infty$ and
$\sigma_{1}^2(\infty)=\infty$.

To determine the asymptotic behavior of $m_{r}^{v}(t)$ as
$t\rightarrow \infty$ we use the formula
\begin{equation}
    D_{-\xi}(0)=2^{\xi/2-1}\frac{\Gamma(\xi/2)}{\Gamma(\xi)},
    \label{eq:WPCF0}
\end{equation}
which follows from Eq.~(\ref{eq:WPCF}). Substituting
Eq.~(\ref{eq:WPCF0}) into Eq.~(\ref{eq:frac_mom}), we obtain
\begin{equation}
    m_{r}^{v}(\infty)=\frac{\Gamma(\xi/2)}{\sqrt{2\pi}}2^{\xi/2}
    \sigma^{\xi-1}(\infty)e^{r^2\sigma_{1}^2(\infty)/2}
    \frac{1+(-1)^v}{2}\quad (\xi>0).
    \label{eq:frac_mom_inf1}
\end{equation}

If Eq.~(\ref{eq:R1}) is fulfilled $[\sigma_{1}^2(\infty)<\infty]$,
all fractional moments (\ref{eq:frac_mom_inf1}) with $r>\alpha-1$
are finite, and the stationary PDF $P_{st}(x)=P_{x}(x,\infty)$ is
given by
\begin{equation}
    P_{st}(x)=\frac{1-\alpha}{2\pi\sigma_{\alpha}(\infty)
    \sigma(\infty)|x|^{\alpha}}\int_{0}^{\infty}du
    \exp\left[-\frac{\ln^{2}(u)}{2\sigma_{\alpha}^{2}(\infty)}
    -\frac{u^2|x|^{2(1-\alpha)}}{2\sigma^{2}(\infty)}\right].
    \label{eq:Pstx1_1}
\end{equation}
Note that $P_{st}(x)$ is even, as is also reflected by
$m_{r}^{1}(\infty)=0$. The value of $P_{st}(x)$ at the point $x=0$
depends on $\alpha$, if $0<\alpha<1$ then $P_{st}(0)=\infty$, if
$\alpha=0$ then
$P_{st}(0)=e^{\sigma_{1}^{2}(\infty)/2}/[\sqrt{2\pi}\sigma(\infty)]$,
and if $\alpha<0$ then $P_{st}(0)=0$.

If Eq.~(\ref{eq:R1}) is not fulfilled
$[\sigma_{1}^2(\infty)=\infty]$, then $m_{r}^{0}(\infty)=\infty$
and $m_{r}^{1}(\infty)=0$. The stationary PDF
$P_{st}(x)=P_{x}(x,\infty)$ (\ref{eq:Px1}) can be written as
\begin{eqnarray}
    P_{st}(x)&\sim&\frac{1}{2\pi\sigma(\infty)|x|^{\alpha}
    \sqrt{2tR_{1}(t)}}  \nonumber\\
    &&\times\int_{0}^{\infty}du
    \exp\left[-\frac{\ln^{2}(u)}{4(1-\alpha)^2tR_{1}(t)}
    -\frac{u^2|x|^{2(1-\alpha)}}{2\sigma^2(\infty)}\right]\;
    (t\rightarrow\infty).
    \label{eq:Pstx1_2}
\end{eqnarray}
For $x\neq 0$, the first term of the exponent in
Eq.~(\ref{eq:Pstx1_2}) can be neglected compared to the second
one, and we find
\begin{equation}
    \left. P_{st}(x)\right|_{x\ne 0}=\lim_{t\to\infty}
    \frac{1}{4|x|\sqrt{\pi tR_{1}(t)}}=0.
    \label{eq:Pstx1_2_1}
\end{equation}
For $x=0$, Eq.~(\ref{eq:Pstx1_2}) is reduced to \linespread{0.5}
\begin{equation}
    P_{st}(0)=\lim_{ \scriptsize\begin{array}{l}
    t\to\infty \\ x\to 0 \end{array} }
    \frac{(1-\alpha)\exp[(1-\alpha)^2tR_{1}(t)]}{\sqrt{2\pi}
    \sigma(\infty)|x|^\alpha}=\infty.
    \label{eq:Pstx1_2_2}
\end{equation}
\linespread{1}

To gain more insight into the behavior of $P_{st}(x)$ in the
vicinity of $x=0$, we define the probability
\begin{equation}
    W_{\varepsilon}(t)=\int_{-\varepsilon}^{\varepsilon}dxP_{x}(x,t),
    \label{eq:W_def}
\end{equation}
that $x(t)\in(-\varepsilon,\varepsilon)$. Substituting
Eq.~(\ref{eq:Pstx1_2}) into Eq.~(\ref{eq:W_def}), we obtain
\begin{equation}
    W_{\varepsilon}(t)\sim\frac{1}{\sqrt{2\pi tA(t)}}
    \int_{-\infty}^{\infty}dz\exp\left[-\frac{z^2}{2tA(t)}\right]
    \mathrm{erf}\left[\frac{\varepsilon^{1-\alpha}e^z}
    {\sqrt{2}\sigma(\infty)}\right]\;(t\rightarrow\infty),
    \label{eq:W1}
\end{equation}
where $A(t)=2(1-\alpha)^2R_{1}(t)$. To evaluate
$W_{\varepsilon}(\infty)$, we introduce the new variable $z=s\sqrt
t$, and take into account that
\begin{equation}
    \lim_{t\rightarrow\infty}\mathrm{erf}
    \left[\frac{\varepsilon^{1-\alpha}e^{s\sqrt t}}
    {\sqrt{2}\sigma(\infty)}\right]=h(s),
    \label{eq:HF}
\end{equation}
where $h(s)$ is the Heaviside function. Substituting
Eq.~(\ref{eq:HF}) into Eq.~(\ref{eq:W1}), we obtain
$W_{\varepsilon}(\infty)=1/2$. Thus, in the case
$\sigma_{1}^2(\infty)=\infty$ no well-defined stationary PDF
exists.

\subsection{$\kappa=0$}

In this subsection, we study the statistical behavior of free
particles (particles not subjected to a restoring force
\cite{DH65}) under influence of two colored noises. As mentioned
in Section 1, free particles driven by additive colored noise or
multiplicative Gaussian colored noise can display anomalous
diffusion and stochastic localization.

In this case ($\kappa=0$) $m(t)=x_{0}^{1-\alpha}$ for all times
$t$, and Eq.~(\ref{eq:disp}) is reduced to
\begin{equation}
    \sigma^{2}(t)=2(1-\alpha)^{2}\int_{0}^{t}du r(u)(t-u).
    \label{eq:disp_k0}
\end{equation}
Eq.~(\ref{eq:disp_k0}) is qualitatively similar to
Eq.~(\ref{eq:disp1}), and its long-time asymptotic has the form
\begin{equation}
    \sigma^{2}(t)\sim 2(1-\alpha)^{2}tR(t)\;(t\rightarrow\infty),
    \label{eq:disp_k0_inf}
\end{equation}
where $R(t)=\int_{0}^{t}du r(u)$. If
\begin{equation}
    R(t)=o(1/t)\;(t\rightarrow \infty),
    \label{eq:R}
\end{equation}
then $\sigma^2(\infty)<\infty$, and if $0<R(\infty)\leq \infty$ or
$R(\infty)=0$ [$R(\infty)$ is the noise intensity], but
Eq.~(\ref{eq:R}) does not hold, then $\sigma^2(\infty)=\infty$.
So, there are two different cases of the long-time asymptotic
behavior of Eq.~(\ref{eq:disp_k0}), namely,
$\sigma^2(\infty)<\infty$ and $\sigma^2(\infty)=\infty$.

It is not difficult to verify that if
$\sigma_{1}^2(\infty)=\infty$ and $\sigma^2(\infty)<\infty$ or
$\sigma^2(\infty)=\infty$ then $m_{r}^{0}(\infty)=\infty$ and
$m_{r}^{1}(\infty)=0$, $P_{x}(0,\infty)=\infty$ and $\left.
P_{x}(x,\infty)\right|_{x\ne 0}=0$, $W_{\varepsilon}(\infty)=1/2$.
So, in this cases no well-defined stationary PDF exists.

If $\sigma^2(\infty)<\infty$ and $\sigma_{1}^2(\infty)<\infty$,
all fractional moments (\ref{eq:frac_mom}) with $r>\alpha-1$ are
finite
\begin{eqnarray}
    m_{r}^{v}(t)&=&\frac{\Gamma(\xi)}{\sqrt{2\pi}}
    \sigma^{\xi-1}(\infty)\exp\left[r^2\sigma_{1}^2(\infty)/2-
    a^2(\infty)/4\right] \nonumber\\
    &&\times\{D_{-\xi}[-a(\infty)]+(-1)^vD_{-\xi}[a(\infty)]\},
    \label{eq:frac_mom_inf3}
\end{eqnarray}
$a(\infty)=x_{0}^{1-\alpha}/\sigma(\infty)$, and the stationary
PDF is given by
\begin{eqnarray}
    P_{st}(x)&=&\frac{1-\alpha}{2\pi\sigma_{1}(\infty)\sigma(\infty)
    |x|^{\alpha}}\int_{0}^{\infty}du\exp\left\{-\frac{\ln^{2}(u)}
    {2\sigma_{\alpha}^{2}(\infty)}-\frac{[ux|x|^{-\alpha}-
    x_{0}^{1-\alpha}]^{2}}{2\sigma^{2}(\infty)}\right\}.\nonumber\\
    \label{eq:Pstx1_3}
\end{eqnarray}
In contrast to the case $\kappa >0$, where the stationary PDF
(\ref{eq:Pstx1_1}) is even, the stationary PDF (\ref{eq:Pstx1_3})
is not even, as is also reflected by the fractional moments,
$m_{r}^{1}(\infty)\neq 0$.

Thus, if Eqs.~(\ref{eq:R1}) and (\ref{eq:R}) are fulfilled, then
Gaussian colored noises $f_{1}(t)$ and $f(t)$ lead to stochastic
localization of free particles. This phenomenon was first
described for free particles driven by additive \cite{DH62} and
multiplicative \cite{DH65} colored noise.

If $\sigma^2(\infty)=\infty$ and $\sigma_{1}^2(\infty)<\infty$,
then $m_{r}^{0}(\infty)=\infty$ and $m_{r}^{1}(\infty)=0$,
$P_{x}(x,\infty)=0$ and $W_{\varepsilon}(\infty)=0$. Consequently,
the stationary PDF does not exist, Gaussian colored noise $f(t)$
gives rise to diffusive behavior of free particles. This case is
qualitatively similar to one considered in Ref. \cite{DH65}. It is
not difficult to verify, if $0<R(\infty)<\infty$, then
$\sigma^2(t)\propto t$ as $t\rightarrow\infty$,
$m_{r}^{0}(t)\propto t^{(\xi-1)/2}$ and $m_{r}^{1}(t)\propto
t^{\xi/2-1}$. The dispersion of the particle position,
$\sigma_{x}^{2}(t)=\langle x^2(t)\rangle-\langle x(t)\rangle^2$,
can be rewritten as
$\sigma_{x}^{2}(t)=m_{2}^{0}(t)-[m_{1}^{1}(t)]^2$. Hence,
$\sigma_{x}^{2}(t)\propto t^{1/(1-\alpha)}$ as
$t\rightarrow\infty$, and the conditions $\alpha=0$, $\alpha<0$,
and $0<\alpha<1$ correspond to normal diffusion, subdiffusion, and
superdiffusion, respectively.

\section{Conclusions}

We have constructed an exactly solvable nonlinear model of an
overdamped system with two multiplicative Gaussian colored noises
and studied its statistical properties. Starting from the exact
solution of the Langevin equation for the case of arbitrary
independent noises, we have considered the specific relation
between noises. In such approach, the time evolution of the system
state has been represented as the ratio of the power-normal
process to the lognormal one, and the two-noise Langevin equation
can be reduced to the one-noise Langevin equation. We have
obtained exact expressions for the time-dependent univariate
probability distribution function of the system state and the
fractional moments. Analyzing their analytically found long-time
asymptotics, we have shown that anomalous diffusion and stochastic
localization of particles, not subjected to a restoring force, can
occur.

\ack

I would like to express my gratitude to S.~I.~Denisov for his
guidance, advice and support.

\end{document}